\documentclass[12pt]{iopart}

\usepackage{iopams}
\usepackage{citesort}
\usepackage{graphicx}
\usepackage{soul,color}
\usepackage{amssymb}

\begin{document}

\title{Time evolution techniques for detectors in relativistic quantum information}

\author{David Edward Bruschi$^{1,2}$, Antony R. Lee$^1$, Ivette Fuentes$^1$\footnote{Previously known as Fuentes-Guridi and Fuentes-Schuller.}}
\address{$^1$School of Mathematical Sciences, 
University of Nottingham, 
Nottingham NG7 2RD, 
United Kingdom}
\address{ $^2$School of Electronic and Electrical Engineering, 
University of Leeds, 
Leeds LS2 9JT, 
United Kingdom}

\begin{abstract}
The techniques employed to solve the interaction of a detector and a quantum field commonly require perturbative methods. We introduce mathematical techniques to solve the time evolution of an arbitrary number of detectors interacting with a quantum field moving in space-time while using non-perturbative methods.  Our techniques apply to harmonic oscillator detectors and can be generalised to treat detectors modelled by quantum fields. Since the interaction Hamiltonian we introduce is quadratic in creation and annihilation operators, we are able to draw from continuous variable techniques commonly employed in quantum optics.
\end{abstract}

\pacs{03.67.-a, 04.62.+v, 42.50.Xa, 42.50.Dv} 

\section{Introduction}
The field of quantum information aims at understanding how to store, process, transmit, and read information efficiently exploiting quantum resources \cite{MikeandIke:00}. 
In the standard quantum information scenarios observers may share entangled states, employ quantum channels, quantum operations, classical resources and perhaps more advanced devices such as quantum memories and quantum computers to achieve their goals.
In order to implement any quantum information protocol, all parties must be able to \textit{locally} manipulate the resources and systems which are being employed. 
Although quantum information has been enormously successful at introducing novel and efficient ways of processing information, it still remains an open question to what extent relativistic effects can be used to enhance current quantum technologies and give rise to new relativistic quantum protocols.

The novel and exciting field of relativistic quantum information has recently gained increasing attention within the scientific community. An important aim of this field is to understand how the state of motion of an observer and gravity affects quantum information tasks. For a review on developments in this direction see \cite{alsing2012}.
Recent work has focussed on developing mathematical techniques to describe localised quantum fields to be used in future relativistic quantum technologies. The systems under investigation include fields confined in moving cavities \cite{alphacentauri2012} and wave-packets \cite{dragan2012,downes2013}. Moving cavities in spacetime can be used to generate observable amounts of bipartite and multipartite entanglement \cite{NFriis:DEBruschi:JLouko:IFuentes:12,PhysRevD.86.105003}. Interestingly, it was shown that the relativistic motion of these systems can be used to implement quantum gates \cite{DEBruschi:Dragan:Lee:JLouko:IFuentes:12}, thus bridging the gap between relativistic-induced effects and quantum information processing. In particular, references \cite{DEBruschi:Dragan:Lee:JLouko:IFuentes:12,PhysRevD.86.105003} employed the covariance matrix formalism within the framework of continuous variables and showed that most of the gates necessary for universal quantum computation could be obtained by simply moving the cavity through especially tailored trajectories \cite{faccio}. This result pioneers on the implementation of quantum gates in Relativistic Quantum Information.

A third local system that has been considered for relativistic quantum information processing is the well known Unruh-DeWitt detector \cite{unruh1976}, a point-like  quantum system which follows a classical trajectory in spacetime and interacts locally with a global free quantum field. Such a system has been employed with different degrees of success in a variety of scenarios, such as  in the work unveiling the celebrated Unruh effect \cite{unruh1976} or to extract entanglement from the vacuum state of a bosonic field \cite{reznik2003}.  Unruh-DeWitt detectors seem convenient for relativistic quantum information processing. However, the mathematical techniques involved, namely perturbation theory, become extremely difficult to handle even for simple quantum information tasks such as teleportation \cite{lin2012}. 

The main aim of our research program is to develop detector models which are mathematically simpler to treat so they can be used in relativistic quantum information tasks. A first step in this direction was taken in \cite{fixandrzej} where a model to treat analytically a finite number of harmonic oscillator detectors interacting with a finite number of modes was proposed exploiting techniques from the theory of continuous variables. The covariance matrix formalism was employed to study the Unruh effect and extraction of entanglement from quantum fields without perturbation theory.  The techniques introduced in \cite{fixandrzej}  are restricted to simple situations in which the time evolution is trivial.  To show in detail how the formalism introduced was applied, the authors presented simplified examples using detectors coupled to a single mode of the field which is formally only applicable when the field can be decomposed into a discrete set of modes with large frequency separation. This situation occurs, for example, when the detectors are inside a cavity. The detector model introduced in this work generalises the model presented in \cite{fixandrzej} to include situations in which the time evolution is non-trivial. 

We introduce the mathematical techniques required to solve the time evolution of a detector, modeled by an harmonic oscillator, which couples to an arbitrary \emph{time-dependent frequency distribution} of modes.  The interaction of the detector with the field is purely quadratic in the operators and, therefore, we can employ the formalism of continuous variables taking advantage of the powerful mathematical techniques that have been developed in the past decade \cite{alsing2012}. These techniques allow us to obtain the explicit time dependent expectation value of relevant observables, such as mean excitation number of particles. As a concrete example, we employ our model to analyse the response of a detector, which moves along an arbitrary trajectory and is coupled to a time-dependent frequency distribution of field modes. 

Recently it was shown that a spatially dependent coupling strength can be engineered to couple a detector to a gaussian distribution of frequency modes \cite{lee2012}. Here we analyse the case where the coupling strength varies in space and time such that the detectors effectively couple to a time evolving frequency distribution of plane waves that can be described by a single mode. A spatial and time dependent coupling strength can be engineered by placing the quantum system in an external potential which is time and space dependent. These tuneable interactions have been produced in ion traps~\cite{thompson1990,miller2005}, cavity QED~\cite{walther1006} and superconducting circuits~\cite{peropadre2010,gambetta2011,srinivasan2011,sabin2012}. In an ion trap, the interaction of the ion with its vibrational modes can be modulated by a time and spatial dependent classical driving field, such as a laser~\cite{haffner2008}. Moreover, in cavity QED, time and space dependent coupling strengths are used to engineer an effective coupling between two cavity modes~\cite{imamoglu1997,guzman2006}.

The techniques we will present simplify the Hamiltonian and an exact time dependent expression for the number operators can be obtained. We also discuss the extent of the impact of the techniques developed in this paper: in particular, we stress that they can be successfully applied for a finite number of detectors following arbitrary trajectories. The formalism is also applicable when the detectors are confined within cavities. In this last case, the complexity of our techniques further simplifies due to the discrete structure of the energy spectrum. Finally, we note that the model can be generalised to the case where the detector is a quantum field itself. 

In this work we adopt the following notation: upper case letters in bold font are used for matrices (i.e. $\boldsymbol{S}$), bold font with subscripts label different matrices (i.e. $\boldsymbol{S}_j$), elements of a matrix $\boldsymbol{S}$ will be printed in plain font with two indices (i.e. $S_{ij}$). Furthermore, vectors of operators appear with blackboard font (i.e. $\mathbb{X}$) and their components by plain font with one index (i.e. $X_i$). Vectors of coordinates are printed in lower case bold font (i.e. $\boldsymbol{x}$) and it will be clear from the context that they differ from the symbols used for matrices.

\section{Interacting systems for relativistic quantum information processing}
Unruh-DeWitt type detectors have been extensively studied in the literature of quantum field theory and relativistic quantum information. In standard quantum field theory, detectors in inertial and accelerated motion have been investigated in \cite{unruh1976,DeWitt1979}. Other investigations looked into different methods of regularising divergent quantities. Such proposals introduced finite interaction time cut--offs and spatial extensions to the detector \cite{higuchi1993,sriramkumar1996,suzuki1997,davies2002,grove1983,takagi1986,schlicht2004,louko2006,satz2007}. In relativistic quantum information Unruh-DeWitt type detectors have been used to create entanglement from the vacuum \cite{reznik2003}, perform quantum teleportation \cite{lin2010}, create past--future entanglement \cite{olson2011,sabin2010}.

A general interaction Hamiltonian $H_I(t)$ between a quantum mechanical system (detector) interacting with a bosonic quantum field $\Phi(t,\mathbf{x})$ in $4$-dimensional~spacetime~is commonly given by
\begin{eqnarray}
H_{I}(t)=\, m(t)\int d^3x \sqrt{-g}\mathcal{F}(t,\mathbf{x})\Phi(t,\mathbf{x}),\label{general:interaction:hamiltonian}
\end{eqnarray}
where $(t,\mathbf{x})$ are a suitable choice of coordinates for the spacetime, $m(t)$ is the monopole moment of the detector and $g$ denotes the determinant of the metric tensor \cite{fabbri2005}. The function $\mathcal{F}(t,\mathbf{x})$ is the effective interaction strength between the detector and the field. When written in momentum space, it describes how the internal degrees of freedom of the detector couple to a time dependent distribution of the field modes. Such details can be determined by a particular physical model of interest. More on the interaction Hamiltonian (\ref{general:interaction:hamiltonian}) can be found in \cite{lee2012,takagi1986,schlicht2004}.

The field $\Phi$ can be expanded in terms of a particular set of solutions to the field equation $\phi_{\mathbf{k}}(t,\mathbf{x})$ as
\begin{eqnarray}
\label{discretedecomp}
\Phi=\sum_{\mathbf{k}}\left[D_{\mathbf{k}}\phi_{\mathbf{k}}+\mbox{h.c.}\right],
\end{eqnarray}
where the variable $\mathbf{k}$ is a set of discrete parameters and $D_{\mathbf{k}}$ are bosonic operators that satisfy the time independent canonical commutation relations $[D_{\mathbf{k}},D_{\mathbf{k'}}^{\dag}]=\delta_{\mathbf{k}\mathbf{k'}}$. We refer to the solutions $\phi_{\mathbf{k}}$ as field modes. We emphasise that the modes $\phi_{\mathbf{k}}$ need \textit{not} be standard solutions to the field equations (i.e. plane waves in the case of a scalar field in Minkowski spacetime) but can also be wave-packets formed by linear superpositions of plane waves.

We can engineer the function $\mathcal{F}(t,\mathbf{x})$ such that 
\begin{eqnarray}
\int d^3x \sqrt{-g}\mathcal{F}(t,\mathbf{x})\Phi(t,\mathbf{x})=h(t)D_{\mathbf{k}_{*}}+\mbox{h.c.},
\end{eqnarray}
where one mode, labelled via $\mathbf{k_{*}}$, has been selected out of the set $\lbrace\phi_{\mathbf{k}}\rbrace$, which in turn implies
\begin{eqnarray}
H_{I}(t)= m(t)\left[h(t)D_{\mathbf{k}_{*}}+\mbox{h.c.}\right].\label{single:mode:interaction:hamiltonian}
\end{eqnarray}
Therefore, the coupling strength has been specially designed to make the detector couple to a single mode,  in this case labeled by $\boldsymbol{k}_{*}$. In the case of a free $1+1$-dimensional relativistic scalar field, the mode the detector couples to corresponds to a time dependent frequency distribution of plane waves. In the following we clarify, using a specific example, what we mean by a time-dependent frequency distribution. The a $1+1$ massless scalar field $\Phi(t,x)$ obeys the standard Klein-Gordon equation $(-\partial_{tt}+\partial_{xx})\phi(t,x)=0$. It can be expanded in terms of standard Minkowski modes (plane waves) as \cite{takagi1986,crispino2008}
\begin{eqnarray}
\Phi(t,x)=\int_{-\infty}^{+\infty} \frac{dk}{\sqrt{2\pi |k|}}\left[a_k e^{-i(|k| t-kx)}+a^{\dagger}_k e^{i(|k| t-kx)}\right],\label{Minkowski:field}
\end{eqnarray}
where the momentum $k\in\mathbb{R}$ and $k>0$ labels right moving modes while $k<0$ labels left moving modes and each particle has energy $\omega:=|k|$. The creation and annihilation operators satisfy the canonical commutation relations $[a_{k},a_{k^{'}}^{\dag}]=\delta(k-k^{'})$. We substitute equation~(\ref{Minkowski:field}) into~(\ref{general:interaction:hamiltonian}), assuming for simplicity a flat spacetime, i.e. $\sqrt{-g}=1$, and by inverting the order of integration we obtain
\begin{eqnarray}
H_{I}(t)=\, m(t)\int_{-\infty}^{+\infty} \frac{dk}{\sqrt{2\pi |k|}}\left[a_k e^{-i|k| t}\tilde{\mathcal{F}}^{*}(t,k)+a^{\dagger}_k e^{i|k|t}\tilde{\mathcal{F}}(t,k)\right]\label{equation:fourier}
\end{eqnarray}
where we have defined the spatial Fourier transform $\tilde{\mathcal{F}}(t,k)$ of the function $\mathcal{F}(t,x)$ as
\begin{eqnarray}
\label{eqn:tdfd}
\tilde{\mathcal{F}}(t,k):=\int_{-\infty}^{+\infty} d^{3}x\mathcal{F}(t,x)e^{-ikx}
\end{eqnarray}
The function ~(\ref{eqn:tdfd}) is the \emph{time dependent frequency distribution}. Thus given a general interaction strength, the momenta contained within the field that interacts with the detector will be modified in a time dependent way.

We should add that our detector model given by equation~(\ref{general:interaction:hamiltonian}) extends the well-known pointlike Unruh-DeWitt detector which has been extensively studied in the literature \cite{DeWitt1979,takagi1986,hu2012}. When the spatial profile approximates a delta function $\mathcal{F}(t,x)=\delta(x(t)-x)$, the detector approximates a point-like system following a classical trajectory $x(t)$ \cite{takagi1986,schlicht2004}. 

In our analysis we have considered the detector to be a harmonic oscillator. By doing this we will be able to draw from continuous variables techniques in quantum optics that will simplify our computations. However, the original Unruh-DeWitt detector consists of a two-level system. The excitation rate of a harmonic oscillator has been shown to approximate well that of a two-level system at short times \cite{lin2010,hu2012}. For long interaction times, the difference becomes significant and the models cannot be compared directly.

In the following, we explain how to solve the time evolution of an arbitrary number of detectors interacting with an arbitrary number of fields when the interaction Hamiltonian is of a purely quadratic form given by equation~(\ref{single:mode:interaction:hamiltonian}).

\section{Time evolution of N interacting bosonic systems}  
We start this section by reviewing from Lie algebra theory and techniques from symplectic geometry. By combining these techniques we will then derive equations that govern the evolution of a quantum system. The generalisation of the quadratic Hamiltonian given by equation (\ref{single:mode:interaction:hamiltonian}) to $N$ interacting bosons is 
\begin{eqnarray} \label{eq:hamil}
H(t)=\sum_{j=1}^{N(2N+1)}\lambda_j(t)G_j,\label{interaction:hamiltonian:expansion}
\end{eqnarray}
where the functions $\lambda_{j}$ are real and the operators $G_{j}$ are Hermitian and quadratic combinations of the harmonic creation and annihilation operators $\lbrace(D_{j},D^{\dagger}_{j})\rbrace$. For example, $G_{1}=D_{1}^{\dag}D_{2}^{\dag}+D_{1}D_{2}$. The summation is over the total number of independent, purely quadratic, operators which for $N$ modes is $N(2N+1)$. The operators $G_i$ form a closed Lie algebra with Lie bracket
\begin{equation}
\label{eqn:structureconstant}
 [G_i,G_j\bigr]=c_{ijk}G_k.
\end{equation}
The algebra generated by the $N(2N+1)$ operators $G_{j}$ is the algebra generated by \textit{all} possible independent quadratic combinations of creation and annihilation bosonic operators. The set of operators $\left\{G_{j}\right\}$ can be divided into four subsets, where $N$ operators generate phase rotations, $2N$ single mode squeezing operations, $N^{2}-N$ independent beam splitting operations and $N^{2}-N$ two mode squeezing operations. Phase rotations and beam splitting together form the well known set of passive transformations \cite{wolf2003}. There are  $(N^{2}-N)+N=N^2$ generators of passive transformations which, excluding the total number operator $\sum D^{\dag}_{i}D_{i}$ that commutes with all passive generators, form the well known sub algebra $SU(N)$ of the total algebra of our model, where $\mathrm{dim}(SU(N))=N^2-1$ \cite{puri1994}.

The complex numbers $c_{ijk}$ are the structure constants of the algebra generated by the operators $G_{j}$. In general they form a tensor that is antisymmetric in its first two indices only. Moreover, the values taken by the $c_{ijk}$ explicitly depend on the choice of representation for the $G_{j}$”.

We wish to find the time evolution of our interacting system. In the general case, the Hamiltonian $H(t)$ does not commute with itself at different times $[H(t),H(t')\bigr]\neq0$. Therefore, the time evolution is induced by the unitary operator
\begin{eqnarray}
U(t)=\overleftarrow{\mathcal{T}}e^{-i\int_{0}^{t} dt' H(t')}\label{evolution:operator}
\end{eqnarray}
where $\overleftarrow{\mathcal{T}}$ stands for the time ordering operator \cite{greiner1996}. We can employ techniques from Lie algebra and symplectic geometry \cite{wilcox1967,berndt2001,hall2004} to explicitly find a solution to equation ~(\ref{evolution:operator}). The unitary evolution of the Hamiltonian can be written as \cite{puri2001},
\begin{eqnarray}
\label{eeo}
U(t)=\prod_{j}U_j(t)=\prod_{j}e^{-iF_j(t)G_j}
\end{eqnarray}
where the functions $F_j(t)$ associated with generators $G_{j}$ are real and depend on time.  By equating (\ref{evolution:operator}) with (\ref{eeo}), differentiating with respect to time and multiplying on the right by $U^{-1}(t)$ we find a sum of similarity transformations
\begin{eqnarray}
H(t)&=&\dot{F}_{1}(t)G_{1}+\dot{F}_{2}(t)U_{1}G_{2}U^{-1}_{1}+\dot{F}_{3}(t)U_{1}U_{2}G_{3}U^{-1}_{2}U^{-1}_{1}+\ldots \label{hamiltonianequation}
\end{eqnarray}
In this way, we obtain a set of $N(2N+1)$ coupled, non-linear, first order ordinary differential equations of the form
\begin{eqnarray}
\label{odes}
\sum_{j}\alpha_{ij}(t)\dot{F}_{j}(t)+\sum_{j}\beta_{ij}F_{j}(t)+\gamma_{i}(t)=0\label{general:ode},
\end{eqnarray}
where the coefficients $\alpha_{ik}(t)$ and $\beta_{ik}(t)$ will in general be functions of the $F_{j}(t)$ and $\lambda_j(t)$. The form of the Hamiltonian and the initial conditions $F_{j}(0)=0$ completely determine the unitary time evolution operator~(\ref{evolution:operator}).

The equations can be re-written in a formalism which simplifies calculations by defining a vector that collects bosonic operators
\begin{eqnarray}
\mathbb{X}:=\left(D_{1},D^{\dagger}_{1},\ldots, D_{N},D_{N}^{\dagger}\right)^T
\end{eqnarray}
In this formalism, successive applications of the Baker-Campbell-Hausdorff formula which are required in the similarity transformations~(\ref{hamiltonianequation}) will be replaced by simple matrix multiplications reducing the problem from a tedious Hilbert space computation to simple linear algebra. We write
\begin{eqnarray}
U_{j}(t)\,G_{k}\,U^{-1}_{j}(t)&=&\mathbb{X}^{\dag}\cdot \mathbf{S}_{j}(t)^{\dag}\cdot \mathbf{G}_{k} \cdot \mathbf{S}_{j}(t) \cdot \mathbb{X}\label{useful:equation}, 
\end{eqnarray}
where we have used the identity $U_{j}(t)\,\mathbb{X}\,U^{-1}_{j}(t)\equiv\mathbf{S}_{j}(t)\cdot\mathbb{X}$ and $\mathbf{G}_{j}$ is the matrix representation of $G_{j}$, defined via $G_{j}:=\mathbb{X}^{\dag}\cdot\mathbf{G}_{j}\cdot\mathbb{X}$. The dynamical transformation of the vector of operators $\mathbb{X}$ generated by the interaction Hamiltonian $G_{j}$ is given by the symplectic matrix \cite{luis1995}
\begin{eqnarray}
\mathbf{S}_{j}:=e^{-iF_{j}(t)\boldsymbol{\Omega}\mathbf{G}_{j}}
\end{eqnarray}
where $F_{j}(t)$ are real functions associated with the generator $G_{j}$ and $\Omega_{ij}:=[X_i,X_j]$ is the symplectic form. A symplectic matrix $\mathbf{S}$ satisfies $\mathbf{S}^{\dag}\,\bOmega\,\mathbf{S}=\bOmega$. In this formalism, we can use~(\ref{useful:equation}) and the identity $H=\mathbb{X}^{\dag}\cdot\mathbf{H}\cdot \mathbb{X}$ to obtain the matrix representation of the Hamiltonian $H$,
\begin{eqnarray}
\label{matrixequations}
\mathbf{H}(t)=&\dot{F}_{1}(t)\,\mathbf{G}_{1}+\dot{F}_{2}(t)\,\mathbf{S}_{1}(t)^{\dag}\cdot\mathbf{G}_{2}\cdot\mathbf{S}_{1}(t)\nonumber\\
&+\dot{F}_{3}(t)\,\mathbf{S}_{1}(t)^{\dag}\cdot\mathbf{S}_{2}(t)^{\dag}\cdot\mathbf{G}_{3}\cdot\mathbf{S}_{2}(t)\cdot\mathbf{S}_{1}(t)+\ldots\label{final:equation}
\end{eqnarray}
It is necessary to explicitly compute the matrix products of the form $\mathbf{S}_{k}(t)^{\dag}\cdot\mathbf{G}_{j}\cdot\mathbf{S}_{k}(t)$ in order to re-write equation (\ref{matrixequations}) in terms of the generators ${\bf G}_i$.
By equating the coefficients of equation (\ref{matrixequations})  to the coefficients $\lambda_j(\tau)$ in the matrix representation of equation (\ref{eq:hamil})
we obtain a set of coupled $N(2N+1)$ ordinary differential equations. Solving for the functions $F_j(t)$, we obtain the explicit expression for the time evolution of the system as described by equation (\ref{eeo}). The final expression is
\begin{eqnarray}
\label{cm:evolution:operator}
\mathbf{S}(t)=\prod_{j}\mathbf{S}_j(t)=\prod_{j}e^{-iF_{j}(t)\boldsymbol{\Omega}\mathbf{G}_{j}},
\end{eqnarray}
which corresponds to the time evolution of the whole system. Systems of great interest are those of Gaussian states which are common in quantum optics and relativistic quantum theory \cite{alsing2012}. In this case the state of the system is encoded by the first moments $\langle X_{j}\rangle$ and a covariance matrix $\boldsymbol{\Gamma}(t)$ defined by 
\begin{eqnarray}
\Gamma_{ij}=\langle X_{i}X_{j}+X_{j}X_{i}\rangle -2\langle X_{i}\rangle\langle X_{j}\rangle.
\end{eqnarray} 
In this formalism, the time evolution of the initial state $\mathbf{\Gamma}(0)$ is given by
\begin{eqnarray}
\label{eqn:symplectictimeevolution}
\mathbf{\Gamma}(t)=\mathbf{S}^{\dag}(t)\mathbf{\Gamma}(0)\mathbf{S}(t).
\end{eqnarray} 

\section{Application: Time evolution of a detector coupled to a field} 
We now apply our formalism to describe a situation of great interest to the field of relativistic quantum information: a single detector following a general trajectory and interacting with a quantum field via a general time and space dependent coupling strength.
We therefore return to our $1+1$ massless scalar field. The standard plane-wave solutions  to the field equation in Minkowski coordinates  are
\begin{equation}
\phi_{k}(t,x)=\frac{1}{2\pi\sqrt{|k|}}e^{-i|k| t+kx}\label{minkowski:modes}
\end{equation} 
which are (Dirac delta) normalized as $ \bigl(\phi_{k},\phi_{k'}\bigr)=\delta(k-k')$ through the standard conserved inner product $\bigl(\cdot,\cdot\bigr)$ \cite{crispino2008}. The mode operators associated with these modes, $a_{k}$, define the Minkowski vacuum via $a_{k}\left| 0\right>_{M}=0$ for all $k$.

The field expansion in equation~(\ref{Minkowski:field}) contains both right and left moving Minkowski plane waves. In general, given an arbitrary trajectory of the detector and an arbitrary interaction strength, the detector couples to both right and left moving waves. However, for the sake of simplicity, in section~\ref{concrete:example:section} we will consider an example where the detector follows an inertial trajectory.  In the 1+1 dimensional case right and left moving waves decouple, therefore it is reasonable to assume that the detector couples only to right moving waves. This situation could correspond to a photodetector which points in one particular direction \cite{downes2013}.

The degrees of freedom of the detector which we assume to be a harmonic oscillator are described by the bosonic operators $d,d^{\dagger}$ that satisfy the usual time independent commutation relations $\bigl[d,d^{\dagger}\bigr]=1$. The vacuum $\bigl|0\bigr>_d$ of the detector is defined by $d\bigl|0\bigr>_d=0$. Therefore, the vacuum $\bigl|0\bigr>$ of the \textit{non-interacting} theory takes the form  $\bigl|0\bigr>:=\bigl|0\bigr>_d\otimes\bigl|0\bigr>_M$. 

In the interaction picture, we use equation~(\ref{general:interaction:hamiltonian}) and assume that the detector couples to the field via the interaction Hamiltonian 
\begin{eqnarray}
H_{I}(t)= m(t)\int dx\, \sqrt{-g}\, \mathcal{F}(t,x)\int_{-\infty}^{+\infty} dk\left[a_{k}\phi_{k}(t,x)+a^{\dagger}_{k}\phi^*_{k}(t,x)\right],\label{example:interaction:hamiltonian}
\end{eqnarray}

Using the Hamiltonian~(\ref{example:interaction:hamiltonian}), we parametrise the interaction via a suitable set of coordinates, $(\tau,\xi)$, that describe a frame comoving with the detector. A standard choice is to use the so-called Fermi-Walker coordinates~\cite{takagi1986,schlicht2004}.
This amounts to expressing $(t,x)$ within the integrals of~(\ref{example:interaction:hamiltonian}) as the functions $(t(\tau,\xi),x(\tau,\xi))$. In the comoving frame, the monopole moment of the detector takes the form
\begin{eqnarray}
m(\tau)=s(\tau)\bigl[e^{-i\Delta \tau}d+e^{i\Delta \tau}d^{\dagger}\bigr],
\end{eqnarray}
where the real function $s(\tau)$ allows us to switch on and off the interaction and $\Delta$ is frequency of the harmonic oscillator. 

In momentum space the detector couples to a time-dependent frequency distribution of Minkowski plane-wave field modes.  In \cite{lee2012} the spatial dependence of the coupling strength was specially designed to couple the detector to peaked distributions of Minkowski or Rindler modes. Here we consider a coupling strength that can be designed to couple the detector to a time-varying wave-packet $\psi(\tau,\xi)$. It is therefore more convenient to decompose the field not in the plane-wave basis but in a special decomposition 
\begin{equation}
\Phi(\tau,\xi)=D_{k_{*}}\psi(\tau,\xi)+D_{k_{*}}^{\dagger}\psi^*(\tau,\xi)+\Phi'(\tau,\xi),\label{discrete:field:decomposition}
\end{equation}
where $\psi$ is the mode the detector couples to, which corresponds to a time dependent frequency distribution of plane waves, and $k_{*}$ represents a particular mode we wish to distinguish in the field expansion. Note that for the case of a field contained within a cavity, where the set of modes is discrete, our methods apply without an explicit need to form discrete wave packets. The operators $D_{k_{*}},D_{k_{*}}^{\dagger}$ are time independent and satisfy the canonical time independent commutation relations $\bigl[D_{k_{*}},D_{k_{*}}^{\dagger}\bigr]=1$. 
The field $\Phi'$ includes all the modes orthogonal to $\psi$ and we will assume them to be countable. Once expressed in the comoving coordinates, the Hamiltonian~(\ref{example:interaction:hamiltonian}) takes on the single mode form ~(\ref{single:mode:interaction:hamiltonian}) when the following conditions are satisfied
\begin{eqnarray}
\label{eqn:singlemodeconditions}
h(\tau) = \int d\xi\,\mathcal{F}(\tau,\xi)\psi(\tau,\xi),\quad
\int d\xi\,\mathcal{F}(\tau,\xi)\Phi'(\tau,\xi) =0\,\,\forall \tau\label{mode:choice:condition}.
\end{eqnarray}
The decomposition~(\ref{discrete:field:decomposition}) can always be formed from a complete orthonormal basis (an example of which can be found in \cite{takagi1986}). In general, the operator $D_{k_{*}}$ does not annihilate the Minkowski vacuum $\bigl|0\bigr>_M$. This observation is a consequence of fundamental ideas that lie at the foundation of quantum field theory, where different and inequivalent definitions of particles can coexist. Such concepts are, for example, at the very core of the Unruh effect~\cite{unruh1976} and the Hawking effect~\cite{hawking1975}. 

The operator $D_{k_{*}}$ will annihilate the vacuum $\bigl|0\bigr>_D$. Note that the vacuum state $\bigl|0\bigr>_I$ of this interacting system is different from the vacuum state $\bigl|0\bigr>$ of the noninteracting theory, i.e. $\bigl|0\bigr>\neq\bigl|0\bigr>_I$. 

Under the conditions~(\ref{eqn:singlemodeconditions}), the interaction Hamiltonian takes a very simple form
\begin{eqnarray}
H_{I}(\tau)=m(\tau)\cdot \left[h(\tau)D_{k_{*}}+h^*(\tau)D_{k_{*}}^{\dagger}\right]\label{interaction:hamiltonian:example}
\end{eqnarray}
which describes the effective interaction between the internal degrees of freedom of a detector following a general trajectory and coupling to a \textit{single mode} of the field described by $D_{k_{*}}$. The time evolution of the system can be solved in this case by employing the techniques we introduced in the previous section.  However, this formalism is directly applicable to describe the interaction of N detectors with the field. In that case, our techniques yield differential equations which can be solved numerically. We choose here to demonstrate our techniques with the single detector case since it is possible to compute a simple expression for the expectation value of the number of particles in the detector. 

Let the detector-field system be in the ground state $\bigl|0\bigr>_D$ at $\tau=0$. We design a coupling such that we obtain an interaction of the form (\ref{interaction:hamiltonian:example}). In this case, the covariance matrix only changes for the detector and our preferred mode. The subsystem described by $d,D_{k_{*}}$ is always separable from the rest of the non-interacting modes. The covariance matrix of the vacuum state $\bigl|0\bigr>_D$ is represented by the $4\times 4$ identity matrix, i.e. $\mathbf{\Gamma}(0)=\mathbf{1}$. From equation~(\ref{eqn:symplectictimeevolution}), the final state $\mathbf{\Gamma}(\tau)$ therefore takes the simple form of $\mathbf{\Gamma}(\tau)=\mathbf{S}^{\dag}\mathbf{S}$. The final state provides the information we need to compute the time dependent expectation value of the detector $N_d(\tau) := \bigl<d^{\dagger}d\bigr>(\tau)$.

From the definition of the covariance matrix $\mathbf{\Gamma}(\tau)$, one finds that $N_d(\tau)$ is related to $\mathbf{\Gamma}(\tau)$ by
\begin{eqnarray}
\label{elements:of:the:state:heisenberg:picture}
\Gamma_{11}(\tau) &=& 2\left<d^{\dagger}d\right>(\tau)-2\left<d^{\dagger}\right>(t)\left<d\right>(\tau)+1
\end{eqnarray}
In this paper we also choose to work with states that have first moments zero, i.e $\langle X_{j}\rangle=0$. In this case, since our interaction is quadratic, the first moments will always remain zero \cite{weedbrook:2012}. Therefore we are left with equation
\begin{eqnarray}\label{elements:of:the:state:heisenberg:picture:final}
\Gamma_{11}(\tau) &=& 2\left<d^{\dagger}d\right>(\tau)+1.
\end{eqnarray}
Our expressions hold for detectors moving along an arbitrary trajectory and coupled to an arbitrary wave-packet. Given a scenario of interest, one can solve the differential equations, obtain the functions $F_{j}(\tau)$ and, by using the decomposition in equation (\ref{cm:evolution:operator}), one can obtain the time evolution of the system. We can find the expression for the average number of excitations in the detector at time $\tau$, which reads
\begin{eqnarray}
\label{eqn:numberexpectation}
N_d(\tau)&=&\frac{1}{2}\left[\mathrm{ch}_{1}\mathrm{ch}_{2}\mathrm{ch}_{3}\mathrm{ch}_{4}-1\right].
\end{eqnarray}
where we have adopted the notation $\mathrm{ch}_{j}\equiv\cosh(2F_{j}(\tau))$. For our choice of initial state we find that the functions $F_{j}(\tau)$ are associated with the generators $G_{1}=d^{\dag}D_{k_{*}}^{\dag}+dD_{k_{*}},G_{2}=-i(dD_{k_{*}}-d^{\dag}D_{k_{*}}^{\dag}),G_{3}=d^{\dag 2}+d^{2}$ and $G_{4}=-i(d^{\dag 2}-d^{2})$,
respectively. The appearance of these functions can be simply related to the physical interpretation of the operators $G_{j}$. In fact, the generators $G_{1}$ and $G_{2}$ are nothing more than the two-mode squeezing operators. Such operations generate entanglement and are known to break particle number conservation. The  two generators $G_{3}$ and $G_{4}$ are related to the single-mode squeezing operators for the mode $d$. The generators $G_{1}\ldots G_{4}$, together with the generators $G_{5}=D_{k_{*}}^{\dag 2}+D_{k_{*}}^{2}$ and $G_{6}=-i(D_{k_{*}}^{\dag 2}-D_{k_{*}}^{2})$ which represent single mode squeezing for the mode $D_{k_{*}}$,  form the set of active transformations of a two mode gaussian state and do not conserve total particle number \cite{arvind1995}. The remaining operators, whose corresponding functions are absent in equation~(\ref{eqn:numberexpectation}), form the passive transformations for gaussian states. These transformations are also known as the generalised beam splitter transformation \cite{luis1995}; they conserve the total particle number of a state and hence do not contribute to equation~(\ref{eqn:numberexpectation}).

It is also of great physical interest to study the response of a detector when the initial state is a different vacua $\bigl|\tilde{0}\bigr>$, for example the Minkowski  vacuum $|0\rangle_{M}$. It is well known \cite{fabbri2005} that the relation between two different vacua, for example $|\tilde{0}\rangle$ and $|0\rangle_{D}$, is
\begin{equation}
\bigl|\tilde{0}\bigr>=N e^{-\frac{1}{2}\sum_{ij}V_{ij}D^{\dagger}_iD^{\dagger}_j}\bigl|0\bigr>_D,\label{vacua:relation}
\end{equation}
where $N$ is a normalisation constant and the symmetric matrix $\mathbf{V}$ is related to the Bogoliubov transformations between two different sets of modes $\{\phi_{k}\}$,$\{\psi_{j}\}$ used in the expansion of the field. The modes $\{\phi_{k}\}$ carry the annihilation operators that annihilate the vacuum $\bigl|\tilde{0}\bigr>$ while the modes $\{\psi_{j}\}$ those that annihilate  the vacuum $\bigl|0\bigr>_D$. In general, the matrix $\mathbf{V}$ takes the form $\mathbf{V}:=\mathbf{B}^*\cdot\mathbf{A}^{-1}$ \cite{fabbri2005}, where the matrices $\mathbf{A}$ and $\mathbf{B}$ collect the Bogoliubov coefficients $A_{jk},B_{jk}$ which for uncharged scalar fields are defined through the inner product $(\cdot,\cdot)$ as
\begin{eqnarray}
A_{jk}=\left(\psi_{j},\phi_{k}\right),\,\,\,\,B_{jk}=-\left(\psi_{j},\phi_{k}^{*}\right).
\end{eqnarray}
In the covariance matrix formalism, we collect the detector operators $d,d^{\dagger}$ and mode operators $D_{i},D_{i}^{\dagger}$ in the vector $\mathbb{X}:=\bigl(d,d^{\dagger},D_{1},D_{1}^{\dagger},D_{2},D_{2}^{\dagger},\ldots\bigr)^T$. The initial state $\mathbf{\Gamma}(0)$ will not be the identity anymore, $\mathbf{\Gamma}(0)=\mathbf{\Theta}\neq\mathbf{1}$.
The final state $\mathbf{\Gamma}(\tau)$ will take the form
\begin{equation}
\boldsymbol{\Gamma}(\tau)=\mathbf{S}^{\dag}\,(\tau)\cdot\mathbf{\Theta}\cdot\mathbf{S}(\tau)
\end{equation}
We assume that within the field expansion using the basis $\{\psi_{j}\}$, the detector interacts only with the mode $\psi_{p}$. We can again apply our techniques to obtain the number expectation value as
\begin{eqnarray}
N_d(t)=&S^{\dagger}_{12}S_{12}+S^{\dagger}_{1(2+p)}S_{1(2+p)}\,\left(\mathbf{V}\cdot\mathbf{V}^{\dagger}\right)_{pp}\nonumber\\
&+S^{\dagger}_{1(3+p)}S_{1(3+p)}\,\left(1+\left(\mathbf{V}\cdot\mathbf{V}^{\dagger}\right)_{pp}\right)\\
&-S^{\dagger}_{1(2+p)}S_{1(3+p)}\,V^*_{pp}-S^{\dagger}_{1(3+p)}S_{1(2+p)}\,V_{pp}.\label{number:expectation:value:bogo}\nonumber
\end{eqnarray}
where $\bigl(\mathbf{V}\cdot\mathbf{V}^{\dagger}\bigr)_{kj}$ denotes the matrix elements of the matrix $\mathbf{V}\cdot\mathbf{V}^{\dagger}$ and we have fixed a field mode labeled by $p$.

\section{Concrete Example: inertial detector interacting with a time-dependent wavepacket\label{concrete:example:section}}
\begin{figure}[t]
\includegraphics[width=0.8\textwidth]{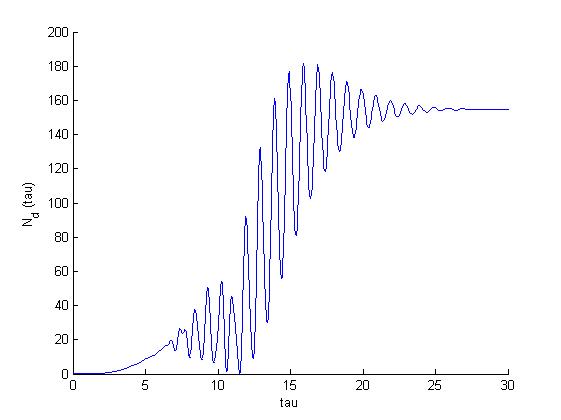}
\centering
\caption{Mean number of particles, $n_{d}(\tau)$, as a function of time $\tau$. Here we used (without loss of generality) $\lambda=1$, $T^2=80$ and $\Delta=2\pi$.\label{figure}}
\end{figure}

To further specify our example we consider the detector stationary and interacting with a localised time dependent frequency distribution of plane waves. The free scalar field is decomposed into wave packets of the form \cite{takagi1986}
\begin{eqnarray}
\tilde{\phi}_{ml}:=\int dk f_{ml}(k)\phi_{k},
\end{eqnarray}
where the distributions $f_{ml}(k)$ are defined as
\begin{equation}
f_{ml}(k):=\left\lbrace\begin{array}{cl} \epsilon^{-1/2}e^{-2i\pi lk/\epsilon} & \epsilon m<k<\epsilon(m+1)\\ 0 & \mathrm{otherwise}\end{array}\right.,
\end{equation}
with $\epsilon>0$ and $\lbrace m,l\rbrace$ running over all integers. Note that if $m\ge 0$ the frequency distribution is composed exclusively of right moving modes defined by $k>0$.  The mode operators associated with these modes are defined as
\begin{eqnarray}
D_{ml}:=\int dk\, f^{*}_{ml}(k)\,a_{k}.
\end{eqnarray}
Notice that for our particular choice of wave-packets, the general operators $D_{j}$ obtain two indices. The distributions $f_{ml}(k)$ satisfy the completeness and orthogonality relations
\begin{equation}
\label{packetrelations}
\sum_{m,l}f_{ml}(k)f^{*}_{ml}(k')=\delta(k-k'),\,\,\,\,\,
\int dk f_{ml}(k)f^{*}_{m'l'}(k)=\delta_{mm'}\delta_{ll'}.
\end{equation}
The wave packets are normalised as $(\tilde{\phi}_{ml},\tilde{\phi}_{m'l'})=\delta_{mm'}\delta_{ll'}$ and the operators satisfy the commutation relations $[D_{ml},D_{m'l'}^{\dag}]=\delta_{mm'}\delta_{ll'}$.
The scalar field can then be expanded in terms of these wave packets as
\begin{equation}
\Phi=\sum_{m,l}\left[\tilde{\phi}_{ml}D_{ml} +h.c.\right].
\end{equation}
Following \cite{schlicht2004}, we consider an inertial detector and we can parametrise our interaction via $t=\tau$ and $x=\xi$. We now construct the spatial profile of the detector to be
\begin{equation}
\label{stationaryprofile}
\mathcal{F}(\tau,\xi):=h(\tau)\int dk f^{*}_{ML}(k)\phi_{k}(\tau,\xi)
\end{equation}
where $h(\tau)$ is now an arbitrary time dependent function which dictates when to switch on and off the detector. Physically, this corresponds to a detector interaction strength that is changing in time to match our preferred mode, labelled by $M,L$ i.e. $D_{k_{*}}=D_{ML}$. We point out that any other wave packet decomposition could be chosen as long as it satisfies completeness and orthogonality relations of the form (\ref{packetrelations}). The form of the spatial profile to pick out these modes is therefore general. Inserting the profile (\ref{stationaryprofile}) into our interaction Hamiltonian (\ref{general:interaction:hamiltonian}), we obtain
\begin{eqnarray}
H_{I}(\tau)=\left(d e^{-i\Delta\tau}+d^{\dag}e^{i\Delta\tau}\right)\left(h(\tau)D_{ML}+h^{*}(\tau)D_{ML}^{\dag}\right)
\end{eqnarray}
We choose the switching on function to be $h(\tau)=\lambda\tau^{2}e^{-\tau^2/T^2}$, where $T$ modulates the interaction time and $\lambda$ quantifies the interaction strength. The interaction Hamiltonian is then
\begin{equation}
H_{I}(\tau)=\lambda\tau^{2}e^{-\frac{\tau^2}{T^2}}\left(d e^{-i\Delta\tau}+d^{\dag}e^{i\Delta\tau}\right)\left(D_{LM}+D_{LM}^{\dag}\right)
\end{equation}
which written in generator form (see equation (\ref{eq:hamil})) is
\begin{eqnarray}
\label{eqn:hamops}
H_{I}(\tau)&=\lambda\tau^{2}e^{-\frac{\tau^2}{T^2}}\left[\cos\left(\tau\Delta\right)G_{1}+\sin\left(\tau\Delta\right)G_{2}\right.\nonumber\\
&\hspace{7mm}+\left.\cos\left(\tau\Delta\right)G_{7}+\sin\left(\tau\Delta\right)G_{8}\right]
\end{eqnarray}
where $G_{7}=d^{\dag}D_{ML}+dD_{ML}^{\dag}$, $G_{8}=-i(dD_{ML}^{\dag}-d^{\dag}D_{ML})$ and the other operators defined similarly. The matrix representation of $H_{I}$ is
\begin{eqnarray}
\label{eqn:hammatrix}
\mathbf{H}_{I}(\tau)=\lambda\frac{\tau^{2}e^{-\frac{\tau^2}{T^2}}}{2}\left[\begin{array}{cccc} 
0 & 0 & e^{i\tau\Delta} & e^{i\tau\Delta} \\
0 & 0 & e^{-i\tau\Delta} & e^{-i\tau\Delta} \\
e^{-i\tau\Delta} & e^{i\tau\Delta} & 0 & 0 \\
e^{-i\tau\Delta} & e^{i\tau\Delta} & 0 & 0
\end{array}\right]
\end{eqnarray}
Equating (\ref{eqn:hammatrix}) and (\ref{matrixequations}), or equivalently (\ref{eqn:hamops}) and (\ref{interaction:hamiltonian:expansion}), gives us the ordinary differential equations we need to find the functions $F_{j}$ for this specific example. Here we solve the equations for $F_{j} (\tau)$ numerically and we plot the average number of detector excitations $n_{d}(\tau)$ as a function of time in figure (\ref{figure}).

We find that the number expectation value of the detector grows and oscillates as a function of time while the detector and field are coupled. This can be expected since the time dependence of the Hamiltonian comes in through complex exponentials that will induce phase rotations in the state and hence oscillations in the number operator. Finally, the number expectation value reaches a constant value after the interaction is turned off. Once the interaction is switched off, the free Hamiltonian does not account for emissions of particles from the detector.

\section{Discussion}
It is of great interest to solve the time evolution of interacting bosonic quantum systems since they are relevant to quantum optics, quantum field theory and relativistic quantum information, among many other research fields. In most cases, it is necessary to employ perturbative techniques which assume a weak coupling between the bosonic systems. In relativistic quantum information, perturbative calculations used to study tasks such as teleportation and extraction of vacuum entanglement \cite{lin2012} become very complicated already for two or three detectors interacting with a quantum field. In cases where the computations become involved, physically motivated or ad hoc approximations can aid, however, in most cases, powerful numerical methods must be invoked and employed to study the time evolution of quantities of interest.

We have provided mathematical methods to derive the differential equations that govern the time evolution of N interacting bosonic modes coupled by a purely quadratic interaction. The techniques we introduce allow for the study of such systems beyond perturbative regimes. The number of coupled differential equations to solve is $N(2N+1)$,  therefore making the problem only polynomially hard.  Symmetries, separable subsets of interacting systems, among other situations can further reduce the number of differential equations. 

The Hamiltonians in our method are applicable to a large class of interactions. In this paper, as a simple example, we have applied our mathematical tools to analyse the time evolution of a single harmonic oscillator detector interacting with a quantum field.  However, our techniques are readily applied to N detectors following any trajectory while interacting with a finite number of wave-packets through an arbitrary interaction strength $\mathcal{F}(t,x)$.  Our techniques simplify greatly when the detectors are confined within a cavity where the field spectrum becomes discrete. The cavity scenario allows one to couple a detector to a single mode of the field in a time independent way as, in principle, no discrete mode decomposition needs to be enforced. Therefore, the single mode interaction Hamiltonian~(\ref{single:mode:interaction:hamiltonian}) can arise in a straightforward fashion.
Inside a cavity, the examples introduced in \cite{fixandrzej} where two harmonic oscillators couple to a single mode of the field are well known to hold trivially.
 
We have further specified our example to analyse the case of an inertial detector interacting with a time-dependent wave-packet. We have showed how to engineer a coupling strength such that the interaction Hamiltonian can be descried by an effective single field mode. However, the field mode is not a plane-wave but a time dependent frequency distribution of plane waves. In this case we have solved the differential equations numerically and showed the number of detector excitations oscillates in time while the detector is on. 
 
Work in progress includes using these detectors to extract field entanglement and perform quantum information tasks. 

\section*{Note}
Near the completion of this work but before posting our results, we became aware of another group working independently along similar lines~\cite{brown2012}. We agreed to post our results simultaneously.

\section{Acknowledgments}
We would like the thank Jorma Louko, Achim Kempf, Sara Tavares, Nicolai Friis and Jandu Dradouma for invaluable discussions and comments.
I. F. acknowledges support from EPSRC (CAF Grant No. EP/G00496X/2).

\section*{References}
\bibliography{refs}

\end{document}